\documentclass[prb,amsfonts,twocolumn]{revtex4}

\def\vec#1{{\bf #1}}

\begin{document}

\title{Pseudopotentials for Multi-particle Interactions in the Quantum Hall Regime}

\author{Steven H. Simon}

\affiliation{Alcatel-Lucent, Bell Labs, Murray Hill, NJ, 07974}

\author{E. H. Rezayi}

\affiliation{Department of Physics, California State University,
 Los Angeles California 90032}

 \author{Nigel R. Cooper}

 \affiliation{T.C.M. Group, Cavendish Laboratory, J.~J.~Thomson Ave., Cambridge, CB3 0HE, United Kingdom}


\begin{abstract}

In fractional quantum Hall physics, the Hilbert space is projected
to a single Landau level and the entire Hamiltonian consists of just
the projected inter-electron interaction.  Haldane's pseudopotential
formalism has been an extremely useful tool both for understanding
these interactions, and for understanding the quantum Hall states
that result. In the current paper we consider the analogue of this
pseudopotential construction that results from general $M$-body
interactions rather than the usual (Coulomb) two-body interaction.
\end{abstract}

\maketitle

\section{Introduction}

In high magnetic fields, the electronic states of two dimensional
systems break up into degenerate bands known as Landau
levels\cite{Prange}. In high enough fields, i.e., in the fractional
quantum Hall regime, one assumes that the spacing between these
levels is large enough such that all of the degrees of freedom are
restricted to a single Landau level. In this case, the entire
Hamiltonian of the system is simply the projected inter-particle
interaction.    In an extremely insightful early paper by
Haldane\cite{Haldane,Prange}, it was pointed out that any
translationally and rotationally invariant two-body interaction,
projected to a single Landau level, could be completely described by
a set of ``pseudopotential coefficients" ($V_L$  with $L \geq 0$)
describing the energy cost to have a pair of particles in a state of
relative angular momentum $L$ (for electrons $L$ must be odd whereas
if we were considering quantum Hall effect of bosons\cite{Bosons}
$L$ would have to be even). This formalism turned out to be useful
not just for describing the details of the interaction, but also for
describing the resulting fractional quantum Hall states.   For
example, the Laughlin $\nu=1/3$ state is precisely described as the
unique highest density zero energy state of a potential which has
$V_1$ positive and $V_L=0$ for all $L>1$. Furthermore, using the
pseudopotential formalism, an interacting system in a partially
filled higher Landau level can be mapped to an equivalent problem in
the lowest Landau level with a modified interaction\cite{Prange}.

In recent years, the study of the fractional quantum Hall effect has
begun to consider interactions beyond just Coulomb two-body
interaction.   Indeed, a particularly interesting class of quantum
Hall states, including the Moore-Read state\cite{MooreRead} and the
Read-Rezayi\cite{ReadRezayi} states  are exact highest density zero
energy states of $M$-body interactions with $M > 2$.  More recent
work\cite{Us1,Us2,Green}has considered even more complicated
$M$-body interactions. These more complicated many particle
interactions may not just be theoretical tools. For example, it has
been shown\cite{CooperExact} how multi-particle interactions may be
engineered in trapped atom systems.   More generally, multi-particle
interactions could easily result from integrating out virtual
excitations of any sort.  Furthermore, while the above mentioned
quantum Hall states are exact ground states for very specific
many-particle interactions, they may also be extremely good trial
states for more realistic interactions\cite{Pfaff}.  More
importantly, in many cases, these exact ground states provide a
particularly simple (and easily studied) representative of an entire
phase of matter.

Despite the increasing interest in these multi-particle
interactions, generalizations of Haldane's formalism have not, to
our knowledge, been systematically constructed.   (Note, however,
that some progress in this direction has been given in
Ref.~\onlinecite{Quinn} in a somewhat different language).
Performing this construction explicitly is the main objective of the
current paper.

The outline of this paper is as follows.  In section
\ref{sec:review} we start with a brief review of Haldane's
pseudopotential construction for simple two-body interactions.  In
Section \ref{sec:Mbody} we consider how this construction should be
generalized to $M$-body interactions. We find that the construction
is more complicated than the two-body case because the simple
pseudopotential coefficients need to be generalized to
pseudopotential matrices. Section \ref{sec:howmany} is devoted to
calculating the dimension of the resulting matrices, and specifying
a convenient basis in which to express these interaction matrices.
In section \ref{sec:conclusions}, we briefly discuss how these
results might be used and why they are interesting.  For clarity,
some simple examples of using these results for three-body
interactions are given in detail in the final appendix.

\section{Review of Haldane's Construction} \label{sec:review}

Before launching into the general construction for $M$-body
interactions, it is useful to review how Haldane's formalism works
for two-body interactions\cite{Haldane,Prange} (albeit in a
slightly different language than that of Haldane's original
construction). Projected to a single Landau level, the Hamiltonian
consists of an interaction term only
\begin{equation}
    H=\sum_{i < j} V(\vec r_i - \vec r_j)
\end{equation}
It will be crucial that the interaction term is both rotationally
and translationally invariant.

Since the Hamiltonian operates on two electrons at a time, we will
consider two electron wavefunctions of the form $\Psi(\vec
r_1,\vec r_2)$. This is sufficient to analyze the effect of the
Hamiltonian in general since any many electron wavefunction can be
decomposed into a sum of a two particle wavefunction times a
wavefunction of everything else.  For example, to find the effect
of the term $V(\vec r_1 - \vec r_2)$ on a general multiparticle
wavefunction $\Psi$, we can always decompose
\begin{equation}
    \Psi(\vec r_1, \ldots, \vec r_N) = \sum_a \Psi_a(\vec
    r_1,
    \vec r_2)  \tilde \Psi_a(\vec r_3, \ldots, \vec r_N)
\end{equation}
and clearly the  $V(\vec r_1 - \vec r_2)$ term of the Hamiltonian
only operates on the $\Psi_a$ term and not the $\tilde \Psi_a$
term.  Thus, we need only focus on two electron wavefunctions
$\Psi(\vec r_i, \vec r_j)$.

It is then useful to further decompose the two electron wavefunction
into center of mass and relative degrees of freedom
\begin{equation}
    \Psi(\vec r_i, \vec r_j) = \sum_{b,c} A_{b,c}   \,   \Psi^{CM}_b\left(\frac{\vec r_i + \vec r_j}{2} \right)
    \Psi^{rel}_c \left(\vec r_i - \vec r_j \right)
\end{equation}
where $\Psi^{CM}_b$ and $\Psi^{rel}_c$ form a basis for the center
of mass and relative wavefunctions respectively.  We note that the
Hamiltonian, which is translationally invariant, acts only on the
relative coordinate.

In the lowest Landau level, using the usual analytic notation for
wavefunctions\cite{Prange} where single particle orbitals are given
by $z^m \exp(-|z|^2/4)$, and the magnetic length is taken to be
unity, we can thus rewrite any product of the wavefunctions
$\Psi^{CM}_b$ and $\Psi^{rel}_c$ as
\begin{equation}
    \psi^{CM}_b\left(\frac{z_i + z_j}{2} \right)
    \psi_c^{rel}\left(z_i - z_j \right) e^{-\frac{1}{4}(|z_1|^2+|z_2|^2)}
\end{equation}
where we have separated out the Gaussian factors.  Both $\psi^{CM}$
and $\psi^{rel}$ must be analytic polynomials, and sometimes we will
notate $\psi^{rel}$ as simply $\psi$. Throughout this paper we will
be discussing the planar (disk) geometry. In Appendix
\ref{app:geometry} other geometries (sphere and torus) are briefly
discussed.

A convenient complete basis to describe the possible relative
wavefunctions $\psi^{rel}$ is now given by
\begin{equation} \label{eq:orthog2particle}
    |L; i,j \rangle = C_{L} \, (z_i - z_j)^{L}
\end{equation}
where $L$ is the relative angular momentum between particles $i$ and
$j$ and $C_L$ is an appropriate normalization constant. Note that if
the particles are fermions then $L$ must be odd, and if they are
bosons, then $L$ must be even. We can now insert this complete set
into the Hamiltonian
\begin{equation}
     H = \sum_{i<j} \sum_{L, L'}  |L; i,j \rangle
     \,\,  \langle L; i,j | V(\vec r_i - \vec r_j) | L'; i,j \rangle
     \,\,  \langle L'; i,j |
\end{equation}
Note that since the interaction is translationally invariant, we
need only insert a complete set for the relative degrees of freedom
(I.e., the interaction is diagonal and trivial in the center of mass
degree of freedom).

The rotational invariance of the interaction now makes the matrix
element diagonal in $L$ so we obtain the Haldane
Hamiltonian\cite{Prange,Haldane}
\begin{equation}
\label{eq:HaldaneHamiltonian}
    H  = \sum_{i<j} \sum_{L}   \,\, V_{L,2} \,\,  P^{L}_{ij}
\end{equation}
where
\begin{equation}
  P^L_{ij} = |L; i,j \rangle \, \langle L; i,j |
\end{equation}
is a projection operator that projects particles $i$ and $j$ to a
state of relative angular momentum $L$ (within the lowest Landau
level).  In Eq.~\ref{eq:HaldaneHamiltonian},
\begin{eqnarray}
    V_{L,2} &=&  \langle L; i,j | V(\vec r_i - \vec r_j) | L; i,j
    \rangle \\&=& \langle L | V | L
    \rangle \label{eq:secondline}
\end{eqnarray}
is known as the pseudopotential coefficient, and we have added the
subscript $2$ here to indicate that this is a two-body interaction.
In Eq.~\ref{eq:secondline} we have written this matrix element in a
convenient shorthand, since by symmetry between particles, $V_{L,2}$
is independent of which $i$ and $j$ is chosen.

One of the great advantages of the pseudopotential formalism is how
easily it generalizes to higher Landau
levels\cite{Chakraborty,Prange} as well as to even more complicated
situations\cite{MacDonald}. The simple generalization stems from the
fact that in any Landau level, there is a one-to-one mapping to
another system completely in the lowest Landau level with a modified
inter-electron interaction\cite{Chakraborty,Prange,MacDonald}.  To
see how this mapping works, we consider a system in the first
excited Landau level (1LL).   Here, one can write any single
electron state as a Landau level raising operator $a^\dagger$
applied to a lowest Landau level state. In this way, an orthogonal
set of relative angular momentum states analogous to
Eq.~\ref{eq:orthog2particle} may be written in the 1LL
\begin{equation}
\label{eq:raise}
    |L; i,j; 1LL \rangle = a^\dagger_i a^\dagger_j |L; i,j \rangle
\end{equation}
where\cite{Chakraborty}
\begin{equation}
\label{eq:raiseop}
    a^\dagger_i = \sqrt{2} (-\partial_{z_i} + z_i^*/4)
\end{equation}
where the derivative acts on the Gaussian factors as well as on the
polynomial part of the wavefunction (and ${}^*$ means complex
conjugation). Using this orthogonal set for states in the 1LL, and
following the same argument, we obtain exactly the same Haldane
Hamiltonian (Eq.~\ref{eq:HaldaneHamiltonian}) but with modified
pseudopotential coefficients
\begin{equation}
    V_{L,2} =  \langle L; i,j |a_i a_j \,\,  V(\vec r_i - \vec r_j) \,\, a^\dagger_i a^\dagger_j | L; i,j
    \rangle.
\end{equation}
Further discussion of Landau level raising (for the case of
three-body interactions) is given in example 4 of appendix
\ref{sec:examples}.

\section{$M$-body interactions} \label{sec:Mbody}

We now generalize Haldane's argument to $M$-body interactions.  We
thus consider a general $M$-body interaction Hamiltonian
\begin{equation}
    H=\sum_{i_1 < i_2 < \ldots < i_M} V(\vec r_1, \ldots, \vec r_M)
\end{equation}
We note that, although not a requirement, we will typically want to
think about $M$-body interactions that are not reducible to sums of
interactions of a smaller number of particles.  This issue is
discussed in more depth in appendix \ref{app:reduce}.

Again crucial to our construction is the assumption that $V$ is both
translationally and rotationally invariant.   We then consider a
general $M$-body wavefunction $\Psi(\vec r_1, \ldots, \vec r_M)$,
which we rewrite in terms of relative and center of mass coordinates
\begin{eqnarray} &&
\Psi(\vec r_1, \ldots, \vec r_M) = \sum_{b,c} A_{b,c} \\ &&
\Psi^{CM}_b\left(\frac{\vec r_1 + \vec r_2 \ldots + \vec r_M}{M}
\right) \Psi^{rel}_c(\vec r_1, \ldots, \vec r_M) \nonumber
\end{eqnarray}
The statement that $\Psi^{rel}$ is a wavefunction for relative
motion means it is translationally invariant, or
\begin{eqnarray} &&
\Psi^{rel}(\vec r_1, \vec r_2, \ldots, \vec r_M)=  \\ &&
~~~~~~~\Psi^{rel}(\vec r_1+\vec x, \vec r_2+\vec x  \ldots, \vec
r_M+\vec x) \nonumber
\end{eqnarray}
for arbitrary $\vec x$.

Once again, we can assume we are in the lowest Landau level so we
can work with analytic forms of the wavefunction, so we write  the
product $\Psi^{CM} \Psi^{rel}$ as
\begin{equation}
\label{eq:sepgauss} \psi^{CM}\left(\mbox{$\frac{1}{M}\sum_{i=1}^M
z_i$}\right) \psi(z_1, \ldots, z_M) e^{-\frac{1}{4} \sum_{i=1}^M
|z_i|^2}
\end{equation}
where $\psi$ is then the relative wavefunction, which must be a
translationally invariant polynomial (and must be antisymmetric for
fermions and symmetric for bosons).  As in the two-body case, we can
categorize these relative wavefunctions in terms of an angular
momentum quantum number. We define the ``relative angular momentum"
operator on the relative wavefunction $\psi$ to be the total degree
of the complex polynomial $\psi$.  If the polynomial $\psi$ is
homogeneous of degree $L$, it is a relative angular momentum
eigenstate with eigenvalue $L$.  Once again we note that our
construction here pertains to the planar (disk) geometry. The sphere
and torus are discussed briefly in appendix \ref{app:geometry}.

As in the two-body case, we are now seeking a complete basis in
which to write our relative wavefunctions.   We can certainly
write any relative wavefunction $\psi$ as a sum of components with
all possible values of $L \geq 0$.  I.e.,
\begin{equation}
 \psi(z_1, \ldots, z_M) = \sum_L \psi_L(z_1, \ldots, z_M)
\end{equation}
where $\psi_L$ has relative angular momentum $L$, i.e, is
homogeneous of degree $L$.  Unfortunately, as compared to the
two-body case, it is no longer true that for a given $L$ there is
only a single possible wavefunction.  Indeed, as we will see in
great detail in section \ref{sec:howmany} below, there may be many
$M$-body wavefunctions with the same relative angular momenta.
Thus, within the space of states with relative angular momentum
$L$, we arbitrarily define an orthonormal basis which we write as
$|L,q\rangle$. The dimension $d$ of this space (and hence the
number of different values of $q$)  depends on $L$, $M$ and
whether we are considering bosons or fermions.   In section
\ref{sec:howmany} we will determine these dependencies.

As in the two-body case, since the interaction is independent of the
center of mass of the $M$ particles, we can then insert this
complete set of relative wavefunctions into the Hamiltonian to
obtain
\begin{equation}
    H=  \!\!\!\!\! \sum_{i_1 < \ldots < i_M}
    \, \sum_{L,q,q'}  |L,q; i_1, \ldots, i_M\rangle
    V_{L,M}^{q,q'}  \langle L,q'; i_1,\ldots, i_M|
\end{equation}
which is a sum over all possible $M$-clusters.  Here the matrix
element is a Hermitian matrix
\begin{eqnarray} \nonumber
V_{L,M}^{q,q'} \!&=&  \langle L,q; i_1,\ldots, i_M| V(\vec r_1,
\ldots ,
\vec r_M)|L,q'; i_1, \ldots, i_M\rangle \\
\! &=&  \langle L,q| V|L,q'\rangle \label{eq:resultnorm}
\end{eqnarray}
where in the second line we have not written the dependence on $i_1,
\ldots, i_M$ since by symmetry again it does not matter which
particles $i_1, \ldots, i_M$ we choose to use to calculate
$V_{L,M}^{q,q'}$.  Again, we use the subscript $M$ here to indicate
that we are considering an $M$-body interaction.    Note that this
matrix element is necessarily diagonal in $L$ by rotational
invariance\cite{endnote1} of the interaction $V$, but is not
generally diagonal in $q$.   For a given $L$, if there are $d$
different basis vectors $|L,q\rangle$, one must then specify the
full $d$ by $d$ Hermitian matrix $V_{L,M}^{q,q'}$.

At this point, it is useful to point out that we need not work
with an orthonormal basis $|L, q \rangle$.  Even given a
nonorthonormal (but linearly independent) basis $|L, r \rangle$ we
can construct an orthonormal $q$-basis as
\begin{equation}
\label{eq:switch}
  |L,q\rangle = \sum_r [R^{-1/2}]_{qr} |L, r \rangle
\end{equation}
where the hermitian matrix $R$ is given by
\begin{equation}
\label{eq:overlaps}
    R_{r,r'} = \langle L,r| L,r' \rangle
\end{equation}
Thus, we can freely translate our Hamiltonian into an arbitrary
$|L,r\rangle$ basis as
\begin{equation}
\label{eq:finalform}
    H=  \!\!\!\!\! \sum_{i_1 < \ldots < i_M}
    \, \sum_{L,r,r'}  |L,r; i_1, \ldots, i_M\rangle
    V_{L,M}^{r,r'}  \langle L,r'; i_1,\ldots, i_M|
\end{equation}
where now
\begin{equation}
\label{eq:resultnonnorm}
    V_{L,M}^{r,r'} = \sum_{r_1,r_1'}  [R^{-1}]_{r'_1,r'} [R^{-1}]_{r,r_1}
    \,\,
 \langle L,r'_1| V|L, r_1 \rangle
\end{equation}
The form of Eq.~\ref{eq:finalform} is generally the analogue of the
Haldane Hamiltonian for $M$-body interactions, and is applicable for
any complete linearly independent basis $|L,r \rangle$ (although it
may in fact be convenient to choose to work with an orthonormal
basis).

While this result seems relatively straightforward, it is somewhat
nontrivial to find a basis $|L,r\rangle$ (or even to compute its
dimension).  This is precisely the task we will undertake in the
next section.

\section{Space of fixed angular momentum relative wavefunctions}
\label{sec:howmany}

We would like to examine the space of $M$-particle translationally
invariant wavefunctions $\psi$ with relative angular momentum $L$.
In other words, we want to study analytic functions in $M$
variables which are homogeneous of degree $L$ and are
antisymmetric for the case of fermions, or symmetric for the case
of bosons.

Let us start by quoting a result to be proven below which applies
directly to the bosonic case:  The dimension of the space of of
translationally invariant symmetric polynomials in $M$ variables
which are homogeneous of degree $L$ is given by
\begin{eqnarray}  \nonumber d_{sym}(L,M) &=& \mbox{number of partitions of the integer $L$ }
\\  \nonumber
&& \mbox{into pieces no larger than $M$ which}  \nonumber \\
&& \mbox{do not include the integer 1} \label{eq:dsym2}
\end{eqnarray}
For example, we have $d_{sym}(7,4)=2$ since there are only two
partitions of $7$ into pieces larger than 1 and no larger than 4
(These partitions are obviously 7=3+4 and 7=3+2+2). A table of
$d_{sym}(L,M)$ is given explicitly for small values of $M$ and $L$
in table \ref{table:dimensions}.   In Appendix \ref{app:partitions}
an explicit analytic form is given for this quantity. The proof of
Eq.~\ref{eq:dsym2} will be performed below by constructing an
explicit basis for these symmetric homogeneous translationally
invariant polynomials.

To handle the fermion case, it is useful to use the result that
any antisymmetric polynomial can be written as a Vandermonde
determinant times a symmetric polynomial\cite{Polynomials}
\begin{equation}
    \psi_{antisym}(z_1 \ldots z_M) = \left[ \prod_{i<j} (z_i - z_j)
    \right]\psi_{sym}(z_1 \ldots z_M)
\end{equation}
Thus, there is a one to one mapping from the space of homogeneous
symmetric polynomials of degree $L = p$ in $M$ variables to the
space of antisymmetric polynomials of degree $L = p + (M(M-1))/2$ in
$M$ variables. It is therefore sufficient to consider only the
symmetric case, and the dimension of antisymmetric homogeneous
functions of degree $L$ in $M$ variables is given by
\begin{equation}
\label{eq:antitosym}
    d_{antisym}(L,M) = d_{sym}(L-M(M-1)/2, M).
\end{equation}

\begin{table}
\begin{tabular}{|c ||c|c|c|c|c|c|c|c|c|c| }
  \hline
\small  $L$ =  & $\,\, 0 \,\,$ & $\,\, 1\,\, $ & $\,\, 2\,\, $ &
$\,\,3\,\,$ & $\,\,4\,\,$ & $\,\,5\,\,$ & $\,\,6\,\,$ &
$\,\,7\,\,$& $\,\,8\,\,$ & $\,\,9\,\,$
 \\
  \hline
  \hline
  $M=2$  & 1 & 0 & 1 & 0 & 1 & 0 & 1 & 0 & 1
  & 0
  \\
  \hline
  $M=3$ &1 &  0 & 1 & 1 &  1 & 1 & 2  & 1 & 2 & 2
  \\
  \hline
  $M=4$ & 1& 0 & 1  & 1 & 2 & 1 & 3 & 2 & 4 & 3
  \\
  \hline
  $M =5$ & 1 & 0 & 1 & 1 & 2 & 2 & 3  & 3 & 5 & 5
  \\
  \hline
 $M=6 $ &1 &  0 & 1 & 1 & 2  & 2  & 4 & 3 & 6 & 6
 \\
    \hline
 $M = 7 $  & 1&  0 & 1 & 1 & 2  & 2  & 4 & 4 & 6 & 7
  \\
 \hline
  $M = 8$  & 1&  0 & 1 & 1 & 2  & 2  & 4 & 4 &  7 & 7
   \\
 \hline
  $M = 9$ & 1 &  0 & 1 & 1 & 2  & 2  & 4 & 4 &  7 & 8
  \\
\hline
\end{tabular}\\
\caption{Table of $d_{sym}(L,M)$, the dimension of the space of
homogeneous symmetric translationally invariant polynomials of
degree $L$ in $M$ variables.   The formula for these entries is
Eq.~\ref{eq:dsym1} or equivalently Eq.~\ref{eq:dsym2}.  The
dimension of the space of homogeneous antisymmetric translationally
invariant polynomials of degree $L$ in $M$ variables is given in
terms of this table by  $d_{antisym}(L,M) = d_{sym}(L-M(M-1)/2,M)$
(see Eq.~\ref{eq:antitosym}).  Note that for an $M$-body
interaction, at angular momentum $L$, there is a $d$ by $d$
Hermitian pseudopotential matrix ($d=d_{sym}(L,M)$ for bosons and
$d_{antisym}(L,M)$ for fermions).} \label{table:dimensions}
\end{table}

We now proceed to construct an explicit basis for the homogeneous
translationally invariant symmetric polynomials of degree $L$ in
$M$ variables.   To do this, we begin with the well known
elementary symmetric polynomials\cite{Polynomials}
\begin{equation}
\label{eq:em}
 e_{m,M}=e_{m,M}(z_1, \ldots, z_M) = \!\!\! \sum_{0< i_1 < i_2 < \ldots i_m \leq M}
\!\!\!  z_{i_1} \ldots z_{i_m}
\end{equation}
for $m \leq M$ and $e_{m,M}$ is defined to be zero for $m > M$. The
products of these elementary polynomials form a linearly independent
basis for all symmetric polynomials in $M$ variables.  (I.e., the
elementary symmetric polynomials generate the mathematical {\it
ring} of symmetric polynomials.)   We can write this basis
explicitly as
\begin{equation}
    v_{\mbox{\scriptsize \boldmath $\lambda$},M} = \prod_{j=1}^M \, [e_{j,M}(z_1,
\ldots, z_M)]^{\lambda_j}
\end{equation}
where {\boldmath$\lambda$} is a vector of nonnegative integers
$\lambda_j$ with $j=1, \ldots, M$.  The degree of the basis vector $
v_{\mbox{\scriptsize \boldmath $\lambda$},M}$ is $L=\sum_{j=1}^M j
\lambda_j$. Thus, we can think of each different
{\boldmath$\lambda$} as representing a partition of the integer $L$
into pieces with each piece no larger than $M$. For example, the
vector $\mbox{{\boldmath$\lambda$}}=(2,1,2,3)$ represents the
partition $(1,1,2,3,3,4,4,4)$ which corresponds to a homogeneous
symmetric polynomial of degree $L=22$.   Thus the dimension of the
space of symmetric homogeneous polynomials of degree $L$ in $M$
variables is given by the number of partitions of the integer $L$
into pieces no larger than $M$ (denoted $P_{L,M}$ in appendix
\ref{app:partitions}).

The subset of polynomials we are interested in are, in addition,
translationally invariant.  Any such polynomial can be written as
a function only of the relative coordinates
\begin{equation}
\label{eq:tildez}
    \tilde z_i = z_i - \frac{1}{M} \sum_{j=1}^M z_j
\end{equation}
We thus consider as our basis polynomials generated by products of $
   e_{m,M}(\tilde z_1, \ldots, \tilde z_M)$
Note however that since $e_{1,M}=e_{1,M}(z_1, \ldots, z_M) =
\sum_{j=1}^M z_j$ we have $e_{1,M}(\tilde z_1, \ldots, \tilde
z_M)=0$.   Thus, our new basis is of the form
\begin{equation}
\label{eq:basisvecs}
    w_{\mbox{\scriptsize \boldmath $\lambda$},M} = \prod_{j=2}^M \, [e_{j,M}(\tilde z_1,
\ldots, \tilde z_M)]^{\lambda_j}
\end{equation}
which for a given {\boldmath $\lambda$} is of degree $L =
\sum_{j=2}^M j \lambda_j$. Analogous to above {\boldmath$\lambda$}
now represents a partition of $L$ into pieces no larger than $M$ but
now none of the pieces are allowed to be 1 (I.e, we can take
$\lambda_1 = 0$). The size of this basis is therefore precisely
given by the rule quoted above in Eq.~\ref{eq:dsym2}.

The $w_{\mbox{\scriptsize \boldmath $\lambda$},M}$ are constructed
to span the space of translationally invariant symmetric
polynomials. However, we must also prove that the all of the
$w_{\mbox{\scriptsize \boldmath $\lambda$},M}$ basis elements are
linearly independent (This is nontrivial since there is a linear
relation between all of the $\tilde z$'s). This proof is more
involved and is relegated to appendix \ref{app:linearindep}.

\section{Discussion}
\label{sec:conclusions}

As in the original Haldane\cite{Haldane} construction of
pseudopotentials for two-body interactions, one of the main uses of
our result is to describe any interaction in a maximally simplified
parameterized form.  However, as with the Haldane case, examining
the form of the pseudopotential coefficients (and in particular,
examining table \ref{table:dimensions}) can have some interesting
ramifications in terms of what type of simple quantum Hall states
might exist. To see this connection, let us revisit the Laughlin
states for a moment. As pointed out by Haldane's original
work\cite{Haldane}, the Laughlin state can be thought of as the
exact highest density zero energy state of an interaction with only
certain nonzero pseudopotentials. For simplicity here, let us
continue to think about quantum Hall effect of bosons (i.e.,
symmetric quantum Hall wavefunctions). If we consider two-body
potentials, examining table \ref{table:dimensions} we see that
$d_{sym}(L,2)=1$ for even $L$ and is zero for odd. Thus there is a
single pseudopotential coefficient for even values of $L$, and no
coefficient for $L$ odd.  I.e., to specify a two-body potential we
specify the coefficients $V_{0,2}, V_{2,2}, V_{4,2}, \ldots$ (The
traditional notation is $V_0$, $V_2$, $V_4$, $\ldots$. Recall the
additional subscript 2 here refers to the fact that we are
considering a two-body interaction). Haldane's interesting
realization\cite{Haldane} is that the Laughlin bosonic $\nu=1/2$
state is the exact highest density zero energy state of a potential
that has $V_{0,2}
> 0$ and all other $V's$ equal to zero. Similarly, if $V_{0,2}>0$ and $V_{2,2}
> 0$, and all other coefficients are zero, the highest density zero energy state is the Laughlin
bosonic $\nu=1/4$ state. These results are, of course, well known.
We emphasize that to obtain these Laughlin states we are simply
forbidding any two particles from having a relative angular momentum
below some cutoff.

Let us now consider three-body interactions for bosons.   Examining
our table, we see that $d_{sym}(L,3)=1$ for $L=0,2,3,4,5,7$.  Thus,
for each of these values of $L$, there is a single pseudopotential
coefficient  (for $L=1$ there is no state, since $d_{sym}(1,3)=0$
and therefore no coefficient). In analogy with Haldane's original
paper, one might naturally ask questions like what is the highest
density zero energy state of a system with $V_{0,3}
> 0$ and all other pseudopotential coefficients equal to zero.  In
other words, what happens if we forbid any three particles from
having zero relative angular momentum.   (Note that this $V_{0,3}$
pseudopotential coefficient also corresponds to a three-body
point-contact interaction, as demonstrated in appendix
\ref{sec:examples}). It is well known that the highest density zero
energy state of such an interaction is the bosonic Moore-Read
Pfaffian\cite{MooreRead,Greiter} at filling fraction $\nu=1$.

We could similarly ask what happens if we make $V_{0,3}
> 0$ and $V_{2,3} > 0$, and all other coefficients zero.
Thus we are asking what happens if we forbid three particles from
having relative angular momentum 0 or 2 (there is no state where
three particles have relative angular momentum 1). The highest
density zero energy state of this interaction, known as the
``Gaffnian", occurs at filling fraction $\nu=2/3$ for bosons and has
been studied in depth in Ref.~\onlinecite{Us1}. (In appendix
\ref{sec:examples} a form of this interaction is given as
derivatives of delta functions). Similarly, if $V_{0,3}$, $V_{2,3}$,
and $V_{3,3}$ are positive and all other coefficients are zero the
highest energy zero energy wavefunction of such an interaction is
known as the Haffnian and has been studied in depth in
Ref.~\onlinecite{Green}. In each case here, we are simply forbidding
three particles from having relative angular momentum below a
certain cutoff and we can consider successively increasing this
cutoff.

One can also generalize this construction to other values of $M$.
For example, the Read-Rezayi state\cite{ReadRezayi} at filling
fraction $\nu=(M-1)/2$ for bosons, is the highest density zero
energy state of a Hamiltonian that consists only of a positive
$V_{0,M}$ coefficient and all other coefficients are zero.  The
generalization of this construction
--- forbidding clusters of $M$ particles to have angular momenum
below some cutoff $L$ --- is studied by the current authors in
Ref.~\onlinecite{Us2}.

When $d_{sym}(L,M) > 1$  there a $d_{sym}$ dimensional
pseudopotential matrix $V_{L,M}^{q,q'}$.  For simplicity, let us
assume that we are working in a basis where the wavefunctions $|L,
q\rangle$ are orthogonal states (recall that each $|L,q\rangle$ is
an $M$-particle relative wavefunction).   If we were to choose the
$V_{L,M}^{q,q'}$ matrix to be positive definite (all eigenvalues
strictly positive), it would forbid (give positive energy to) any
cluster of $M$ particles with relative angular momentum $L$.  This
is the situation that is considered in Ref.~\onlinecite{Us2}.

However, we could also consider more interesting situations. If we
choose $V_{L,M}^{q,q'}$ to have some eigenvectors with positive
eigenvalue and some eigenvectors with zero eigenvalue, we allow only
certain clusters of $M$ particles with angular momentum $L$ while
disallowing (giving positive energy to) others. To be more specific,
any wavefunction describing a cluster of $M$-particles that lies
completely in the space of zero eigenvalue eigenvectors (the ``null
space") is a zero energy cluster, whereas any cluster that has
components outside of this null-space has positive energy. (An
explicit example of this is given in Appendix \ref{sec:examples}).

 It is
particularly interesting to consider the situation where the matrix
$V_{L,M}^{q,q'}$ has a single eigenvector $|\phi\rangle$ with zero
eigenvalue (and all other eigenvalues are positive).  What this
means is that in the $d_{sym}(L,M)$ dimensional space of
translationally invariant homogeneous polynomials of degree $L$ in
$M$-variables, there is only a single (translationally invariant)
polynomial $\phi(z_1, \ldots, z_M)$ that is a zero energy
$M$-particle wavefunction, and all other polynomials are positive
energy wavefunctions.  We may then ask whether a quantum Hall
wavefunction exists with the property that all clusters of $M$
particles are in the relative wavefunction $\phi$. If such a
wavefunction does exist, it would certainly be a zero energy state
of our chosen Hamiltonian. In some cases it may also be the highest
density zero energy state, i.e., the ground state of the
Hamiltonian.  In future work, we will show a number of examples
where this is indeed the case.

Such a general construction could be quite significant, as it allows
us to construct a different quantum Hall wavefunction for each
(translationally invariant) cluster function $\phi$ that we might
define. In cases where the dimension $d_{sym}(L,M)= 1$, this is no
freedom at all -- there is only a single possible cluster function
$\phi$. However, when $d_{sym} > 1$, we have a continuum of possible
choices --- and therefore can define a continuum of possible
Hamiltonians with a continuum of ground states wavefunctions.

Most interestingly, by judiciously choosing the cluster function
$\phi$ we may be able to find Hamiltonians whose ground state are
quantum Hall wavefunctions corresponding to known unitary conformal
field theories (For discussion of the connection between quantum
hall and conformal field theory see, Refs.
\onlinecite{MooreRead,ReadRezayi,TopologicalReview}). Without proof
we believe such constructions are generically possible, and we will
show several examples of this in future work.  The general belief is
that any wavefunction corresponding to a unitary conformal field
theory corresponds to a stable phase of matter.

To summarize our work, the main purpose of this paper was to
formulate Haldane pseudopotentials for multi-particle interactions.
The main technical result of this paper is the enumeration of the
possible relative wavefunctions for clusters of $M$-particles. Since
we only consider interactions that are rotationally invariant, we
categorize the possible $M$-particle translationally invariant
cluster wavefunctions by their total angular momentum. For bosons,
we find $d_{sym}(L,M)$ basis states whereas for fermions we
calculate $d_{antisym}(L,M)$.   We discuss how a rotationally
invariant $M$-body interaction can be decomposed into
pseudopotential coefficient matrices $V_{L,M}^{q,q'}$.  For a given
$L$ and $M$ the matrix is $d$ by $d$ (with $d$ being $d_{sym}$ for
bosons and $d_{antisym}$ for fermions. We further describe a
relatively simple linearly independent (but not orthonormal) basis
in which such an interaction matrix could be expressed.

For increased clarity, we would like to show the some simple
examples of using this pseudopotential formalism.  In particular, we
will give details for bosonic systems with three-body potentials.
However, since these examples are somewhat lengthy we have put them
in Appendix \ref{sec:examples}.  We encourage the reader to work
through the calculations given there to cement the general ideas of
this paper. A brief summary of the appendix is as follows:

\begin{itemize}
\item We write down the $w_{\mbox{\scriptsize \boldmath $\lambda$},3}$
basis for three-cluster wavefunctions (as discussed in section
\ref{sec:howmany} above)
\item Since this basis is not orthonormal, we show how
to orthonormalize (as discussed in section \ref{sec:Mbody})
\item We consider several simple toy model three-body interactions
and write down their pseudopotential coefficients in this basis (as
discussed in section \ref{sec:Mbody})
\item We show how to calculate pseudopotential coefficients
for higher Landau levels (analogous to that discussed in section
\ref{sec:review})
\end{itemize}

\noindent The explicit demonstration of these steps in the appendix
should make concrete the abstract ideas expressed above.

\vspace*{10pt}

 {\bf Acknowledgements:}  This work was partially
supported by the UK EPSRC Grant No.  GR/S61263/01 and the ICAM
senior fellowship Programme (N. R. C).   E.H.R.~acknowledges support
from DOE under contract DE-FG03-02ER-45981. All three authors
acknowledge the hospitality of the KITP in Santa Barbara where some
of this work was initiated. Conversations with F.~D.~M.~Haldane are
gratefully acknowledged.

\appendix

\vspace*{10pt}

\section{Sphere and Torus Geometries}
\label{app:geometry}

On the spherical geometry\cite{Haldane}, angular momentum is a good
quantum number, and any particle in the LLL has angular momentum $S
=N_{\phi}/2$ with $N_\phi$ the total number of flux quanta through
the sphere. The maximum total angular momentum of $M$ particles is
$M S$ (if they are bosons).  What we called ``relative angular
momentum" $L$ on the plane corresponds to total angular momentum $M
S - L$ on the sphere.   For fermions, since the minimum relative
angular momentum is $M(M-1)/2$, the maximum total angular momentum
is $M S - M(M-1)/2$.  The direction of the ``total" angular momentum
plays the role of center of mass coordinate on the plane.
Interactions that are rotationally invariant, again, will not mix
states of differing ``relative" angular momentum.

A way to think about the translation from plane to sphere is to
realize that any $M$-particle cluster wavefunction is completely
determined by its behavior as all of the particles come close
together --- the ``analyticity" of the wavefunction implies that we
know the complete wavefunction over the entire system if we know its
behavior for short distances. Thus, if we look at $M$ particles in
some relative wavefunction very close together, they do not ``see"
the curvature of the sphere --- locally we can think of the
particles as living on a plane, and so our understanding of the
possible wavefunctions on the plane can tell us all the possible
wavefunctions on the sphere. Once we specify the analytic form of
the wavefunction for the particles very close together, this
uniquely specifies the wavefunction over the entire sphere, thus
generating an appropriate basis for ``relative" wavefunctions on the
sphere.  In other words, for each of our relative wavefunction
states on the plane, there is exactly one relative wavefunction on
the sphere with the same analytic form at short distances.

In fact, it turns out that there is an extremely simple translation
from the plane to sphere.   If we point our center of mass
coordinate towards the north pole, there is then a simple mapping of
the planar wavefunctions (with center of mass at position zero) to
the sphere, by stereographic projection\cite{ReadRezayi}. The only
slight complication is that the usual gaussian factors on the plane
become modified
\begin{equation}
e^{-|z|^2/4} \rightarrow \frac{1}{(1 + |z|^2/4R^2)^{1+S}}
\end{equation}
where $R$ is the radius of the sphere.

On the torus, the story is somewhat more complicated, but in
principle similar.  First of all, on the torus angular momentum is
not conserved (for arbitrary unit cell). Nonetheless, we can define
a linearly independent basis for relative wavefunctions.  As with
the above discussion for the sphere when we bring the particles
close together, the systems looks locally just like an infinite
plane. For each state in our linearly independent basis for the
plane there is exactly one state on the torus that has the same
short distance limit.  Thus, in principle, we can convert our basis
on the plane into a basis on the torus.

\section{Reducible Interactions}
\label{app:reduce}

Given an $M$-body interaction $V(\vec r_1, \ldots, \vec r_M)$  it is
frequently convenient to separate out any part of an interaction
that is ``reducible" to interactions between any lower number $M' <
M$ of particles.  For example, a three-body interaction of the form
\begin{equation}
V(\vec r_1, \vec r_2, \vec r_3) = V_2(\vec r_1 - \vec r_2)+ V_2(\vec
r_2 - \vec r_3) +  V_2(\vec r_1 - \vec r_3)
\end{equation}
is fully reducible to two-body components.

For an arbitrary $M$-body interaction, it is easy to find the
$M'$-body reducible component (with $M' < M$)  by taking appropriate
limits where $M-M'$ of the particles are moved away to infinity. For
example, for a three-body interaction, the two-body part is given by
\begin{equation}
V_2(\vec r_1 - \vec r_2) =\lim_{\vec r_3 \rightarrow \infty} V(\vec
r_1, \vec r_2, \vec r_3)
\end{equation}
We then have the irreducible three-body interaction given by
\begin{eqnarray}
& & ~~~~~~~~ V_{irr}(\vec r_1, \vec r_2, \vec r_3) = V(\vec r_1,
\vec r_2, \vec r_3)  \\  &-& [ V_2(\vec r_1 - \vec r_2) + V_2(\vec
r_2 - \vec r_3) + V_2(\vec r_1 - \vec r_3) ] \nonumber
\end{eqnarray}
We note that it is possible (although unnecessarily complicated) to
describe any two-body interaction as an unreduced three-body
interaction.  However, typically it is simpler (and clearer) to
fully reduce three-body interactions and treat their two-body piece
separately.  For most realistic potentials, the two-body piece will
be very large compared to the reduced three-body piece (which in
turn will be large compared to the reduced four-body piece, and so
forth).  However, by fine-tuning, in certain situations it can be
arranged that the pure two-body piece vanishes\cite{CooperExact}.

\section{Analtyic Expression for Partitions}

\label{app:partitions}

We claim that Eq.~\ref{eq:dsym2} can be reexpressed as
\begin{equation}
\label{eq:dsym1} d_{sym}(L,M) = P_{L,M} - P_{L-1,M}
\end{equation}
where $P_{L,M}$ is the number of partitions of the integer $L$
into pieces no larger than $M$. (Note also that $P_{L,M}$ is
equivalently the number of ways to partition the integer $L$ into
no more than $M$ pieces).

To see the equivalence between Eq.~\ref{eq:dsym1} and
Eq.~\ref{eq:dsym2} we simply note that given any partition of $L$
into pieces no larger than $M$ which includes the integer $1$, we
can drop this $1$ and (uniquely) generate a partition of $L-1$ into
pieces no larger than $M$.     A generating function for $P_{L,M}$
is given analytically as
\begin{equation}
  Z_M(q)=  \prod_{m=1}^M \frac{1}{1 - q^m} = \sum_{L=1}^\infty  \,   q^L P_{L,M}
\end{equation}
so that
\begin{equation}
P_{L,M} = \left[ \frac{1}{L!} \left( \frac{d}{dq} \right)^L Z_M(q)
\right]_{q=0}
\end{equation}

\section{Proof of Linear Independence}
\label{app:linearindep}

Linear independence of the $w_{\mbox{\scriptsize \boldmath
$\lambda$},M}$ basis can be reformulated as follows: Given a
polynomial $P$ in the space spanned by the $w_{\mbox{\scriptsize
\boldmath $\lambda$},M}$ (i.e, given a translationally invariant
symmetric polynomial in $M$ variables) this polynomial can be
written as
\begin{equation}
P=\sum_{\mbox{\boldmath $\lambda$}} \,\,  a_{\mbox{\scriptsize \boldmath
$\lambda$}} \,\,  w_{\mbox{\scriptsize \boldmath $ \lambda$},M}
\label{eq:unique}
\end{equation}
where $\lambda_1 = 0$ in all terms of the sum (since the
$w_{\mbox{\scriptsize \boldmath $\lambda$},M}$ basis spans the
space). Proof of linear independence of all of the
$w_{\mbox{\scriptsize \boldmath $\lambda$},M}$ basis vectors is
equivalent to the statement that each such polynomial corresponds to
a unique set of coefficients
 $a_{\mbox{\scriptsize \boldmath $\lambda$}}$.

To prove this linear independence, we make use of translational
invariance again and shift the center of mass coordinate of all $M$
of the $z_i$'s to zero. Thus, we can rewrite Eq.~\ref{eq:basisvecs}
as
\begin{equation}
\label{eq:tildev}    w_{\mbox{\scriptsize \boldmath $ \lambda$},M} =
\prod_{j=2}^M \, [e_{j,M}(z_1, \ldots, z_{M-1}, \, \underline{z}\,
)]^{\lambda_j}
\end{equation}
where
\begin{equation}
    \underline{z} = \, -\sum_{j=1}^{M-1} \, z_j
\end{equation}
Here, $w_{\mbox{\scriptsize \boldmath $ \lambda$},M}$  must be a
symmetric polynomial in the remaining $M-1$ variables $z_j$. Indeed,
it is easy to check that
\begin{eqnarray} \nonumber
& &    e_{m,M}(z_1, \ldots, z_{M-1},\underline{z}) = e_{m,M-1}(z_1,
\ldots, z_{M-1})   \nonumber
\\  \nonumber
 &-& e_{m-1,M-1}(z_1, \ldots, z_{M-1}) \,\, e_{1,M-1}(z_1,
\ldots, z_{M-1}) \\ &=& e_{m,M-1} - \,  e_{m-1,M-1} \,\, e_{1,M-1}
\label{eq:emm}
\end{eqnarray}
where again $e_{m,M-1}$ is defined to be zero for $m \geq M$. In
particular note that $e_{M,M-1} =0$ and
\begin{equation}
    e_{M,M}(z_1, \ldots, z_{M-1}, \, \underline{z}\, ) =
e_{M-1,M-1}e_{1,M-1} \label{eq:emm2}
\end{equation}
By rewriting our basis vectors $w_{\mbox{\scriptsize \boldmath
$\lambda$},M}$ in terms of the $e_{m,M-1}$ we will be able to use
the fact that the basis vectors generated by these $e_{m,M-1}$
(i.e., the $v_{\mbox{\scriptsize \boldmath $\lambda$},M-1}$ basis)
are linearly independent.

We proceed as follows:   Given the polynomial $P$ written in terms
of the $w_{\mbox{\scriptsize \boldmath $ \lambda$}}$ basis vectors
as in Eq.~\ref{eq:unique}, we rewrite each $w_{\mbox{\scriptsize
\boldmath $ \lambda$}}$ as sums of products of $e_{m,M-1}$ using
Eq.~\ref{eq:emm}.  I.e., we are rewriting everything in terms of the
$v_{\mbox{\scriptsize \boldmath $\lambda$},M-1}$ basis.  We then
group terms such that  $P=P_0 + P_0'$ where $P_0$ includes terms
with no factors of $e_{1,M-1}$ and $P_0'$ includes terms with at
least one factor of $e_{1,M-1}$.  Writing  $P_0$ in the
$v_{\mbox{\scriptsize \boldmath $\lambda$},M-1}$ basis we have
\begin{equation} P_0=\sum_{\mbox{\scriptsize \boldmath
$\lambda$}^{(0)}} \,\, a^{(0)}_{\mbox{\scriptsize \boldmath
$\lambda$}^{(0)}} \,\, v_{\mbox{\scriptsize \boldmath
$\lambda$}^{(0)},M-1}
\end{equation}
where since $P_0$ has no factors of $e_{1,M-1}$, the vectors
$\mbox{\boldmath $ \lambda$}^{(0)}$ must have $\lambda_1^{(0)} =0$,
and we must also have $ \lambda_m^{(0)} = 0$ for $m \geq M$ since
$e_{m,M-1} = 0$ for $m \geq M$. We then construct,
\begin{equation}
Q_0 = \sum_{\mbox{\scriptsize \boldmath $\lambda$}^{(0)}} \,\,
a^{(0)}_{\mbox{\scriptsize \boldmath $\lambda$}^{(0)}} \,\,
w_{\mbox{\scriptsize \boldmath $\lambda$}^{(0)},M}
\end{equation}
with the basis $w_{\mbox{\scriptsize \boldmath $\lambda$},M}$
defined as in Eq.~\ref{eq:tildev}.   Examining Eq.~\ref{eq:emm}, we
see that the $e_{m,M-1}$ terms generate precisely the polynomial
$P_0$ within the expansion of $Q_0$.  The other terms in the
expansion of $Q_0$ all contain $e_{1,M-1}$.   Note that unlike
$P_0$, we have constructed $Q_0$ to be in the space spanned by
$w_{\mbox{\scriptsize \boldmath $\lambda$},M}$ (i.e., it is a
translationally invariant polynomial in $M$ variables). Thus, since
$P$ is in the space spanned by $w_{\mbox{\scriptsize \boldmath
$\lambda$},M}$ then $P - Q_0$ is also in the space spanned by
$w_{\mbox{\scriptsize \boldmath $\lambda$},M}$.  The difference
$P-Q_0$ is constructed so as to have $e_{1,M-1}$ in all of its
terms. Examining Eq.~\ref{eq:emm} and Eq.~\ref{eq:emm2} we see that
if a polynomial ($P-Q_0$ in this case) is in the space spanned by
$w_{\mbox{\scriptsize \boldmath $\lambda$},M}$ and it has
$e_{1,M-1}$ in all of its terms, it must also contain a factor of
$e_{M-1,M-1}$ (i.e., this factor could only have come from a factor
of $e_{M,M}$ in $w_{\mbox{\scriptsize \boldmath $\lambda$},M}$) or
it must be zero. Thus, we can write
\begin{eqnarray}
    P &=& Q_0 - e_{1,M} e_{M-1,M-1} P_1 \\
    &=& Q_0 + e_{M,M}(z_1, \ldots, z_{M-1}, \underline{z})
    \, P_1
\end{eqnarray}
and now $P_1$ is a polynomial of degree less than that of $P$ and is
in the space spanned by $w_{\mbox{\scriptsize \boldmath $
\lambda$},M}$.  We then iterate this procedure, to rewrite $P_1$ as
a sum of two terms, $Q_1$ containing no factors of $e_{1,M-1}$ and
another term $e_{M,M} P_2$ with $P_2$ of lower degree still, and so
forth.  Eventually for some $\alpha_{max}$, we must obtain
$P_{\alpha_{max}} - Q_{\alpha_{max}} = 0$, which then terminates the
procedure. Thus, we can successively decompose into a finite sum
\begin{equation}
\label{eq:penult}
    P=\sum_{\alpha=0}^{\alpha_{\max}} \,\, [e_{M,M}(z_1, \ldots, z_{M-1}, \underline{z})]^\alpha  \,\,
    Q_\alpha
\end{equation}
Where at each level of the iteration, we have
\begin{equation}
\label{eq:ult} Q_\alpha = \sum_{\mbox{\scriptsize \boldmath
$\lambda$}^{(\alpha)}} \,\,  a^{(\alpha)}_{\mbox{\scriptsize
\boldmath $\lambda$}^{(\alpha)}} \,\, w_{\mbox{\scriptsize \boldmath
$\lambda$}^{(\alpha)},M}
\end{equation}
where the vectors $\mbox{\boldmath $ \lambda$}^{(\alpha)}$ must have
$\lambda_1^{(\alpha)} = \lambda_M^{(\alpha)} = 0$. We realize that
Eqs. \ref{eq:penult} and \ref{eq:ult} can be recast in the form of
Eq.~\ref{eq:unique} where the coefficients $a_{\mbox{\scriptsize
\boldmath $\lambda$}}$ are determined by $a_{\mbox{\scriptsize
\boldmath $\lambda$}} = a^{(\alpha)}_{\mbox{\scriptsize \boldmath
$\lambda$}^{(\alpha)}}$ when $\lambda_M = \alpha$ and
$\lambda_k^{(\alpha)} = \lambda_k$ for all $k \neq M$ (and we always
have $\lambda^{(\alpha)}_1 =
 \lambda_1 = 0$, and $\lambda^{(\alpha)}_m = \lambda_m =0$ for $m >
M$).

Our construction here is unique in the sense that given any
polynomial $P$ in the space spanned by  $w_{\mbox{\scriptsize
\boldmath $\lambda$},M}$ we find a unique set of coefficients
$a_{\mbox{\scriptsize \boldmath $\lambda$}}$ in Eq.~\ref{eq:unique}.
This then completes our proof of the linear independence of the
basis $w_{\mbox{\scriptsize \boldmath $\lambda$},M}$.

\section{Examples: Bosons with three-body interactions}
\label{sec:examples}

To elucidate our results, we will  explicitly consider bosons
interating via three-body interactions.   As discussed at length
above, we will be concerned with three-body wavefunctions.   In
calculating any matrix element of these wavefunctions, we mean
\begin{equation}
 \langle \Psi_1 | \hat O | \Psi_2 \rangle = \int d\mu \,  [\Psi_1(z_1,z_2,z_3)]^* \, \hat O
 \, \Psi_2(z_1,z_2,z_3)
\end{equation}
where $\hat O$ is any operator and the integration measure is
\begin{equation}
    d\mu = dz_1 \, dz_1^* \, dz_2 \, dz_2^* \, dz_3 \, dz_3^*
\end{equation}
Recall that our full wavefunctions include both the translationally
invariant relative wavefunction $\psi$ as well as and the center of
mass coordinate $\Psi^{CM}$, and the Gaussian factors (See
Eq.~\ref{eq:sepgauss})
\begin{equation}
    \Psi_i = \psi^{CM}_i\left(\frac{z_1 + z_2 + z_3}{3}\right)  \,\,  \psi_{i}(z_1,z_2,z_3)
    \,\, e^{-\frac{1}{4}\sum_{i=1}^3 |z_i|^2}
\end{equation}
Assuming that $\hat O$ is translationally invariant, it only couples
to $\psi^{rel}$.   Thus, we can choose an orthogonal basis for
$\psi^{CM}$ and we will have the matrix element be zero if
$\psi^{CM}_1$ is orthogonal to $\psi^{CM}_2$.  If we choose
$\psi^{CM}_1 = \psi^{CM}_2$ then, since the operator is
translationally invariant, the matrix element should be independent
of our different possible choice of $\psi^{CM}$, so long as the
different possible choices are normalized the same way. For
performing calculations, it is then acceptable to choose, for
example, $\psi^{CM}_1=\psi^{CM}_2=1$, and we can then check that our
results are independent of this choice.  This also implies a
normalization choice for our matrix elements. We will work with this
choice throughout this appendix.

For the relative wavefunctions, we will work in the
$w_{\mbox{\scriptsize \boldmath $\lambda$},M}$ basis discussed
above.    We write out the first few three-cluster wavefunctions
here explicitly\cite{endnoteket}
\begin{eqnarray}
\label{eq:basisarray}
|L=0\rangle &=& w_{(0,0,0),3} = 1\\
|L=2\rangle &=& w_{(0,1,0),3} = e_{2,3}(\tilde z_1, \tilde z_2,
\tilde z_3)\nonumber
  \\
|L=3\rangle &=& w_{(0,0,1),3} = e_{3,3}(\tilde z_1, \tilde z_2,
\tilde z_3) \nonumber
 \\
|L=4\rangle &=& w_{(0,2,0),3} = [e_{2,3}(\tilde z_1, \tilde z_2,
\tilde z_3)]^2 \nonumber
 \\
|L=5\rangle &=& w_{(0,1,1),3} = e_{2,3}(\tilde z_1, \tilde z_2,
\tilde z_3) e_{3,3}(\tilde z_1, \tilde z_2, \tilde z_3) \nonumber \\
|L=6,a\rangle &=& w_{(0,3,0),3} = [e_{2,2}(\tilde z_1, \tilde z_2,
\tilde z_3)]^3\nonumber \\
|L=6,b\rangle &=& w_{(0,0,2),3} = [e_{2,3}(\tilde z_1, \tilde z_2,
\tilde z_3)]^2 \nonumber\\
|L=7\rangle &=& w_{(0,2,1),3} = [e_{2,2}(\tilde z_1, \tilde z_2,
\tilde z_3)]^2 e_{3,3}(\tilde z_1, \tilde z_2, \tilde z_3) \nonumber
\\ &\vdots& \nonumber
\end{eqnarray}
and recall the definition, Eq \ref{eq:tildez}, of $\tilde z_i$ in
terms of $z_i$.   Note that none of these basis states are
normalized.  Further, note that since $d_{sym}(L=6,3) = 2$ there are
two basis states for $L=6$.   The two states, as discussed above,
are linearly independent, but not orthogonal (or normalized).  We
remind the reader that a wavefunction with relative angular momentum
$L$ vanishes as $L$ powers when all three particles come to the same
point.

We note in passing that for the particular case of three body
wavefunctions, another basis has been constructed that is already
orthonormalized\cite{Laughlin}.   However, this construction does
not generalize easily to more than three particles, so we choose to
work with the above basis to show the more general method.

We start by calculating the normalization of these basis states, by
setting the operator $\hat O=1$.  The integrations over $z_i$'s can
then be performed straightforwardly (Mathematica makes this trivial)
to give the normalizations
\begin{eqnarray}
\label{eq:normalizationarray}
\langle L=0 | L=0 \rangle &=& 2^3 \pi^3 \\
\nonumber\langle L=2 | L=2 \rangle &=& 2^5 \pi^3 \\
\nonumber\langle L=3 | L=3 \rangle &=& 2^8 \pi^3/3^2 \\
\nonumber\langle L=4 | L=4 \rangle &=& 2^9 \pi^3 \\
\nonumber\langle L=5 | L=5 \rangle &=& 2^{12}\pi^3/3^2 \\
\nonumber\langle L=7 | L=7 \rangle &=& 5*2^{15} \pi^3/3^2
\end{eqnarray}
and for $L=6$ the full overlap matrix (See Eq.~\ref{eq:overlaps}) is
given by
\begin{equation}
\label{eq:normalizationmatrix}
\left(\begin{array}{cc} \! \langle a | a \rangle &  \langle a | b \rangle \\
\! \langle b | a \rangle & \langle b | b \rangle
\end{array} \right) = \pi^3  \left(\begin{array}{cc} 3^2*2^{11} &  -2^{12}/3 \\
-2^{12}/3  & ~~~~11*2^{13}/3^4
\end{array} \right)
\end{equation}

{\bf Example 1:} As an example we now consider the toy model
three-body point contact interaction
\begin{equation}\label{eq:threepoint}
    V(\vec r_1, \vec r_2, \vec r_3) = \tilde V \, \delta(\vec r_1 -
    \vec r_2)\, \delta(\vec r_2 -
    \vec r_3)
\end{equation}
Note that this three-body interaction is fully reduced in the sense
of Appendix \ref{app:reduce}. Allowing the delta functions to act,
the interaction matrix element then reduces to
\begin{equation}
\label{eq:deltares} \langle \Psi_1 | V | \Psi_2 \rangle  = \tilde V
\!\! \int \!\!\!dz dz^* [\psi_1(z,z,z)]^*  \,  \psi_2(z,z,z) e^{-3
|z|^2/2}
\end{equation}
From our basis states, only the state $|L=0\rangle=1$ does not
vanish when all three particles come to the same position, so the
only nonzero matrix element is trivially calculated to be
\begin{equation}
\label{eq:deltamatrix0}
    \langle L=0|V|L=0\rangle =  (2 \pi/3)\tilde V
\end{equation}
Thus, the only nonzero pseudopotential coefficient is $V_{0,3}$.  In
our non-normalized basis (See Eq.~\ref{eq:resultnonnorm}) we obtain
\begin{equation}
    V_{0,3} =  \frac{\langle L=0 | V | L=0 \rangle}{|\langle L=0 |
    L=0 \rangle|^2} = \frac{\tilde V}{3 (2 \pi)^5}
\end{equation}
Alternately, we could construct a normalized basis (See
Eq.~\ref{eq:switch})
\begin{equation}
    |L=0\rangle_{Norm} = |L=0\rangle/ \sqrt{8 \pi^3}
\end{equation}
in terms of which we would have (See \ref{eq:resultnorm}) a
pseudopotential coefficient
\begin{equation}
    V_{0,3} =  {}_{Norm}\langle L=0 | V | L=0 \rangle_{Norm} = \frac{\tilde V}{3 (2 \pi)^2}
\end{equation}
As mentioned above, the highest density zero energy state of this
interaction is precisely the Moore-Read Pfaffian
state\cite{MooreRead,Greiter}.

{\bf Example 2:} Let us now consider a more complicated interaction.
Consider
\begin{equation}
    V(\vec r_1, \vec r_2, \vec r_3) = \tilde V \,\, \nabla_1^4 \, [\, \delta(\vec r_1 -
    \vec r_2)\, \delta(\vec r_2 -
    \vec r_3)\, ]
\end{equation}
where the subscript $1$ on $\nabla^4$ means we are differentiating
with respect to the position $\vec r_1$.   Derivative of delta
function interactions, such as this one, are quite useful since
arbitrary interactions can be built up as a series of successive
derivatives of delta functions (We can think of such a series as a
Taylor series expansion in fourier space).

To handle this interaction, we first integrate by parts then let one
delta function act to give
\begin{eqnarray}
& & \langle \Psi_1 | V | \Psi_2 \rangle  =  \tilde V \int d\vec r_1
d\vec r_2   \delta(\vec r_1 - \vec r_2) e^{-|\vec r_2|^2} \times \nonumber \\
& & \nabla^4_1\ \left[ [\psi_1(z_1,z_2, z_2)]^* \psi_2( z_1, z_2,
z_2) e^{-|\vec z_1|^2/2} \right] \label{eq:sln}
\end{eqnarray}
Using the fact that
\begin{equation}
\label{eq:delsquare} \nabla^2= 4 \partial_z
\partial_{z^*}
\end{equation}
 the second line of Eq.~\ref{eq:sln} then becomes
\begin{eqnarray}
& &  (4^2) \left[(\partial_{z_1^*} - z_1/2)^2 \nonumber
[\psi_1(z_1,z_2,z_2)]^*\right] \times \\ & & \left[ (\partial_{z_1}
- z_1^*/2)^2 \psi_2(z_1,z_2,z_2) \right] e^{-|z_1|^2/2}
\end{eqnarray}
Once the last delta function acts, all three of the particles are
put at the same position.   As we mentioned above, the wavefunction
for a state with angular momentum $L$ vanishes as $L$ powers when
all the particles come to the same position.   Now, since we have up
to two derivatives, we see that both $L=0$ and $L=2$ can have
nonzero matrix elements, but no higher $L$ wavefunction can.  The
$L=0$ matrix element is messy to calculate (although it is easy on
Mathematica) and gives
\begin{equation}
\langle L=0|V|L=0\rangle =   (4^3 \pi/27)\tilde V
\end{equation}
The $L=2$ matrix element, on the other hand, is actually easy to
calculate.  Here, we only need to keep the terms where all of the
derivatives act on the wavefunctions (the $z/2$ and $z^*/2$ do not
contribute).  With our wavefunction being $e_{2,3}$, we note that
\begin{eqnarray} \nonumber
\partial_{z_1} e_{2,3}(\tilde z_1,\tilde z_2, \tilde z_3) &=& \partial_{\tilde z_1} e_{2,3}(\tilde z_1,\tilde z_2, \tilde z_3)
= (\tilde z_2 + \tilde z_3) \\ \label{eq:use1}
\partial_{z_1}^2 e_{2,3}(\tilde z_1,\tilde z_2, \tilde z_3) &=& -2/3
\end{eqnarray}
so that
\begin{equation}
\label{eq:de2}
\partial_{z_1}^2 |L=2\rangle \, = (-2/3) \, |L=0\rangle
\end{equation}
Thus the matrix element is given by $(-2/3)^2$ times the value of
Eq.~\ref{eq:deltamatrix0} times the prefactor of $4^2$ from
Eq.~\ref{eq:delsquare}
\begin{equation}
\langle L=2|V|L=2\rangle = (4^2)(8 \pi/27) \tilde V
\end{equation}
Again, we could write this in either the normalized, or unnormalized
basis to give the pseudopotential coefficient $V_{2,3}$.

{\bf Example 3:} Let us now try a much more complicated interaction.
Consider
\begin{equation}
    V(\vec r_1, \vec r_2, \vec r_3) = \tilde V \, \nabla_1^{12}\,  [\, \delta(\vec r_1 -
    \vec r_2)\, \delta(\vec r_2 -
    \vec r_3) \, ]
    \label{eq:12deriv}
\end{equation}
The reason we choose this is because it has nontrivial matrix
elements of $L=6$, which is the first ``interesting" case where
there are two states at the same $L$.   As above, we begin by
integrating by parts.  As in Example 2 above, calculation of matrix
elements for $L< 6$ is quite messy, but for $L=6$ things simplify
quite a bit.  Analgous to the $L=2$ case for Example 2 above, the
only term that does not vanish when the delta function acts is the
one where all of the derivatives have been applied. Noting that from
Eq.~\ref{eq:basisarray} we have $|L=6,a\rangle = [e_{2,3}]^3$ and
$|L=6,b\rangle =[e_{3,3}]^2$ we then have
\begin{eqnarray} \nonumber
\!\!\!\!\! \alpha &\equiv&    \partial_{z_1}^6 [e_{2,3}]^3  =
\frac{6!}{2!2!2!}
   [ \partial_{z_1}^2 e_{2,3}]^3 = -80/3 \\
\!\!\!\!\! \beta &\equiv& \partial_{z_1}^6  [e_{3,3}]^2 =
\frac{6!}{3!3!}
   [ \partial_{z_1}^3 e_{3,3}]^2 = 320/81 \nonumber
\end{eqnarray}
Where we have used Eq.~\ref{eq:use1} above, as well as the analogous
easily calculated $\partial_{z_1}^3 e_{3,3} = 4/9$.   With these
facts, to determine the matrix elements of $V$ we realize that we
have just the same integral as in Eq.~\ref{eq:deltamatrix0} above
times these factors (and a prefactor of $4^6$ coming from
Eq.~\ref{eq:delsquare})
\begin{equation} \label{eq:Vmatrixform}
\left(\begin{array}{cc} \! \langle a |V| a \rangle &  \langle a |V| b \rangle \\
\! \langle b |V| a \rangle & \langle b |V| b \rangle
\end{array} \right) = (4^6)(2 \pi/3)\tilde V \left(\begin{array}{cc} \alpha^2 &  \alpha \beta \\
\alpha \beta   & \beta^2
\end{array} \right)
\end{equation}
Had we chosen to consider instead
\begin{equation}
    V(\vec r_1, \vec r_2, \vec r_3) = \tilde V\, \nabla_1^{6} \nabla_2^6 \, [\, \delta(\vec r_1 -
    \vec r_2)\, \delta(\vec r_2 -
    \vec r_3)\, ] \label{eq:66deriv}
\end{equation}
we would have obtained the same form of Eq.~\ref{eq:Vmatrixform} but
with
\begin{eqnarray}
\alpha &\equiv&    \partial_{z_1}^3 \partial_{z_2}^3 [e_{2,3}]^3 =
28/3 \\
\beta &\equiv& \partial_{z_1}^3 \partial_{z_2}^3 [e_{3,3}]^2  =
104/81
\end{eqnarray}
Or analogously we might have chosen the interaction
\begin{equation}
    V(\vec r_1, \vec r_2, \vec r_3) = \tilde V \, \nabla_1^{4} \nabla_2^4 \nabla_3^4 \, [\, \delta(\vec r_1 -
    \vec r_2)\delta(\vec r_2 -
    \vec r_3) \, ]
\end{equation}
to obtain the form of Eq.~\ref{eq:Vmatrixform} but with
\begin{eqnarray}
\alpha &\equiv&    \partial_{z_1}^2 \partial_{z_2}^2
\partial_{z_3}^2 [e_{2,3}]^3 =
-8/3 \\
\beta &\equiv& \partial_{z_1}^2 \partial_{z_2}^2
\partial_{z_3}^2 [e_{3,3}]^2 = 176/81
\end{eqnarray}
and so forth (obviously there are many more possibilities in the
same spirit).

It is interesting to note that independent of which of these
interactions we choose, the form of Eq.~\ref{eq:Vmatrixform} is rank
one --- meaning that it leaves some three-cluster wavefunctions with
$L=6$ at zero energy as discussed in section \ref{sec:conclusions}.
If we want to give positive energy to all three-cluster relative
states with $L=6$, we can do so by adding together two of these
interactions, for example adding the interaction in
Eq.~\ref{eq:12deriv} to the interaction in Eq.~\ref{eq:66deriv} (and
we should choose $\tilde V$ positive in both cases).

Given the result in Eq.~\ref{eq:Vmatrixform}, and given the
normalization matrix Eq.~\ref{eq:normalizationmatrix}, we can plug
into Eq.~\ref{eq:resultnonnorm} to obtain the pseudopotential matrix
$V_{6,3}^{r,r'}$ in this non-orthonormal basis.  Or, we could
construct an orthonormal basis using Eq.~\ref{eq:switch}, and then
construct the pseudopotential matrix $V_{6,3}$ in this basis as in
Eq.~\ref{eq:resultnorm}.

{\bf Example 4:} Let us now consider bosons in the first excited
Landau level interacting with a three-body potential. Note that this
is a bit of an artificial problem since most interesting boson
problems would be lowest Landau level. Nonetheless, it is a well
defined question of what the pseudopotentials would be for bosons in
any given Landau level. Analogous to Eq.~\ref{eq:raise}, we can
raise any three-particle wavefunction into the first excited Landau
level by applying $a^\dagger_1 a^\dagger_2 a^\dagger_3$ to the
wavefunction, with the raising operator given by
Eq.~\ref{eq:raiseop}. To raise to the $n^{th}$ excited Landau level,
we would use $(a^\dagger)^n$.

We can thus raise the entire basis given in Eq.~\ref{eq:basisarray}
to create a basis in the first excited Landau level, which we write
as
\begin{eqnarray} \nonumber
    |L=0\rangle_{1LL} &=& a^\dagger_1 a^\dagger_2
a^\dagger_3     |L=0\rangle  = z^*_1 z^*_2 z^*_3 / 2 \sqrt{2} \\
    |L=2\rangle_{1LL} &=& a^\dagger_1 a^\dagger_2
a^\dagger_3     |L=2\rangle  = \ldots \nonumber \\
    &\vdots&
\end{eqnarray}
with the unraised states on the right defined as in
Eq.~\ref{eq:basisarray}.  We have written out the case of $L=0$
explicitly (without the Gaussian factor), but the case of $L=2$ is a
rather long expression.  It is worth noting, however, that very
generally, the raised function $|L \rangle_{1LL}$ does not vanish
when all three particles come to the same position for $L \leq 3$
(the number 3 occurs here because there are 3 particles, therefore 3
raising operators, therefore a maximum of 3 derivatives).

It is convenient that the normalizations of these states are
unchanged
\begin{eqnarray}
 {}_{1LL}\langle L=0 | L=0
\rangle_{1LL} &=& \langle L=0 | L=0 \rangle \\
\nonumber_{1LL}\langle L=2 | L=2 \rangle_{1LL} &=& \langle L=2 | L=2
\rangle \\
& \vdots& \nonumber
\end{eqnarray}
since $[a^\dagger, a]=1$, and LLL states plays the role of the
$a^\dagger$ vacuum.

Let us now consider the simple three point delta function
interaction of Eq.~\ref{eq:threepoint}.  Allowing the delta
functions to act we obtain the matrix elements as a single remaining
integral as in Eq.~\ref{eq:deltares}.  It is trivial to analytically
obtain the result for $L=0$ and $L=3$ (we leave this as an exercise
for the reader). For $L=2$, however, the calculation is done with
the aid of Mathematica.   We end up with
\begin{eqnarray}
{}_{1LL}\langle L=0 | V| L=0 \rangle_{1LL} &=& (4 \pi/27) \tilde V \\
{}_{1LL}\langle L=2 | V| L=2 \rangle_{1LL} &=& (8 \pi/9) \tilde V  \nonumber \\
{}_{1LL}\langle L=3 | V| L=3 \rangle_{1LL} &=& (256 \pi/243) \tilde
 V \nonumber
\end{eqnarray}
and matrix elements for higher $L$ all vanish.


\begin{thebibliography}{22}
\bibitem{endnote1}  To see more clearly that the matrix elements are diagonal in $L$,
consider the behavior of the matrix element under the rotation
generated by $z \rightarrow e^{i \theta} z$. If a rotationally
invariant $V$ is sandwiched between any two homogeneous
polynomials of the same degree, the phase factors will cancel.
However, if $V$ is sandwiched between two polynomials of differing
degree (differing $L$) the resulting matrix element will depend on
$e^{i \theta}$ which implies either the matrix element is
nonrotationally invariant, or it vanishes. Since we have assumed
both $V$ and the wavefunctions are rotationally invariant, the
only possibility is that the matrix element between wavefunctions
of differing $L$'s vanishes.
\bibitem{endnoteket} Here we have a slight abuse of notation: the
left hand side of Eq. \ref{eq:basisarray} is a ``ket" vector whereas
the right hand side is the explicit analytic part of the coordinate
representation.
\end{thebibliography}
\end{document}